\documentclass[usenatbib]{basi}

\usepackage[T1]{fontenc}
\usepackage[british]{babel}
\usepackage[varg]{txfonts}

\newcommand{\bs}{Benchmark  Star}
\newcommand{\feh}{[Fe/H]}
\newcommand{\teff}{$\mathrm{T}_{\mathrm{eff}}$}
\newcommand{\logg}{$\log g$}
\newcommand{\vmic}{$v_{\mathrm{mic}}$}

\newcommand{\alfCen}{$\alpha$ Cen}
\newcommand{\alfCet}{$\alpha$ Cet}
\newcommand{\alfTau}{$\alpha$ Tau}
\newcommand{\betAra}{$\beta$ Ara}
\newcommand{\betGem}{$\beta$ Gem}
\newcommand{\betHyi}{$\beta$ Hyi}
\newcommand{\betVir}{$\beta$ Vir}
\newcommand{\delEri}{$\delta$ Eri}
\newcommand{\epsEri}{$\epsilon$ Eri}
\newcommand{\epsFor}{$\epsilon$ For}
\newcommand{\epsVir}{$\epsilon$ Vir}
\newcommand{\etaBoo}{$\eta$ Boo}
\newcommand{\gamSge}{$\gamma$ Sge}
\newcommand{\ksiHya}{$\xi$ Hya}
\newcommand{\muAra}{$\mu$ Ara}
\newcommand{\muCas}{$\mu$ Cas}
\newcommand{\muLeo}{$\mu$ Leo}
\newcommand{\psiPhe}{$\psi$ Phe}
\newcommand{\tauCet}{$\tau$ Cet}
\newcommand{\cyg}{61~Cyg~}

\newcommand{\tab}[1]{Table~\ref{#1}}
\newcommand{\fig}[1]{Fig.~\ref{#1}}
\newcommand{\sect}[1]{Sect.~\ref{#1}}

\newcommand{\FeI}{Fe~{\small I}}
\newcommand{\FeII}{Fe~{\small II}}

\begin{document}

\title[Gaia Benchmark  Stars]{Gaia FGK Benchmark  Stars and their reference parameters}

\author[Jofr\'e et al.]{Paula Jofr\'e$^{1,2}$ \thanks{email: \texttt{pjofre@ast.cam.ac.uk}}, 
    Ulrike Heiter$^3$\thanks{email: \texttt{ulrike.heiter@physics.uu.se}},
    Sergi Blanco-Cuaresma$^2$ and Caroline Soubiran$^2$
    \\
    $^1$ Institute of Astronomy, University of Cambridge, Madingley Road, Cambridge CB3 0HA, U.K. \\
    $^2$ Laboratoire d'Astrophysique de Bordeaux (UMR 5804),  Univ. Bordeaux - CNRS, F-33270, Floirac, France.\\
    $^3$ Department of Physics and Astronomy,  Uppsala University, Box 516, 75120 Uppsala, Sweden}

\pubyear{2012}
\volume{00}
\pagerange{\pageref{firstpage}--\pageref{lastpage}}

\date{Received --- ; accepted ---}

\maketitle

\label{firstpage}

\begin{abstract}
In this article we summarise on-going work on the so-called Gaia FGK Benchmark  Stars. This work consists of the determination of their atmospheric parameters and  of the construction of a high-resolution spectral library. The definition of such a set of reference stars has become crucial in the current era of large spectroscopic surveys. Only with homogeneous and well documented stellar parameters can one exploit these surveys consistently and understand the structure and history of the Milky Way and therefore other of galaxies in the Universe. 
\end{abstract}


\section{Introduction}\label{sec:intro}

Stellar spectral libraries are commonly needed for two immediate purposes: (1) to build population synthesis models, which help us to understand the structure and evolution of galaxies; (2) to evaluate methods to determine stellar atmospheric parameters, which help us to understand the structure and evolution of stars and thus the Milky Way. Spectral libraries can be built from observations or from theoretical models. Hence, for a thorough comprehension of stellar spectra, accurate atomic and molecular data as well as atmospheric models are necessary. Additionally good quality observations are required to validate the modelled spectra.

The Sun has been so far the (benchmark) star most widely used to calibrate and evaluate analyses of stellar spectra. The Sun represents, however, only a fraction of spectral-types  (G-type) of stars in our Galaxy.  To understand how to develop atmospheric models, parametrisation pipelines of stellar spectroscopic surveys, and therefore proper stellar spectral libraries, we need more benchmark  stars representing FGK stars of various metallicities which are so important for galactic studies. 



Our aim is to define such a set of benchmark  stars and provide homogeneous parameters for them whose determination is well documented. These stars have been chosen several years ago to be the pillars of the calibration of the parameters that will be derived for one billion stars by Gaia. Observations on the NARVAL instrument in France and HARPS instrument in Chile have been conducted in the past years for that purpose. We encourage to use these stars as part of any strategy to calibrate, evaluate, and homogenise different methods and databases, such as the Gaia-ESO Survey  \citep{2012Msngr.147...25G}  is doing with its current $\sim$15000 observations. This work on the Gaia Benchmark  Stars will be fully described in three different articles. The first one (Heiter et al. 2013, Paper~I) explains the selection criteria of \bs s and the determination of the effective temperature (\teff) and surface gravity (\logg). The second one (Blanco-Cuaresma et al. 2013, Paper~II) introduces  our spectral librarie of \bs s. The third one \citep[Paper~III]{2013arXiv1309.1099J} consists of a detailed spectral analysis of this library in order to define a metallicity scale (\feh)  for these stars.

Here we present a summary of this work, we also discuss the implications of the uncertainties and finally we give a general comparison of our parameters to previous parameters found in the literature.

 \section{Determination of Atmospheric Parameters}\label{params_sect}
The key aspect of the Gaia Benchmark  Stars is that the stellar parameters \teff\ and \logg\ are determined using fundamental relations, that means,  independently from the spectra. For a star to be one of our benchmarks, we need to know, a priori, its radius, its bolometric flux and its distance. Additionally, it must be bright enough to obtain very high signal-to-noise and high resolution spectra.  Our first set of stars consists of 34 Hipparcos stars and covers a wide range of stellar parameters. The list of stars with their basic properties, together with their atmospheric parameters is in \tab{params}.

 \subsection{Effective temperature}
The effective temperature is determined from the Stefan-Boltzmann relation 
\begin{equation}
\mathrm{F_{bol}}= \sigma (0.5 \theta_{\mathrm{LD}})^2 \mathrm{T_{eff}}^4
\end{equation}
where $\mathrm{F_{bol}}$ is the bolometric flux, $\theta_{\mathrm{LD}}$ is the angular diameter of the star, and $\sigma$ is the Stefan-Bolzmann constant.  About 70\% of the stars have a direct measurement of their radius via interferometry, while the rest has radii using calibrations, such as infrared spectrophotometry and photometric surface-brightness relations. The bolometric flux is also determined only for  half of the stars directly from the integration of the flux over the whole spectrum. For the rest, photometric relations are used. Uncertainties of the angular diameter and the bolometric flux are taken into account in the error of the temperature.

 \subsection{Surface Gravity}
The surface gravity is determined by the Newton's law of gravity  
\begin{equation}
g = \frac{GM}  {(0.5 \theta_{\mathrm{LD}}/\pi)^2}
\end{equation} 
where $G$ is the gravitational constant and $M$ and $\pi$ are the mass and parallax of the star, respectively.  To determine the mass, we considered stellar evolution models using the luminosity derived from the bolometric flux and the parallax, the direct effective temperature and initially a  metallicity value from the literature. If our estimated metallicity (see below) is significantly different from the chosen literature value, we  re-estimated \logg\ using our own result. We used  
different evolutionary tracks to estimate an error for the mass. Uncertainties in the angular diameter, the mass and the parallax are taken into account for the error in \logg.   More details about the direct stellar parameter measurements will be found in the forthcoming Paper~I.

 \subsection{Metallicity}\label{met}
 The metallicity was determined by analysing  \FeI\ and \FeII\ lines  in high resolution (R = 70000) and high signal-to-noise spectra. 
  Iron abundances were estimated by fixing \teff\ and \logg\ to their fundamental values. 
   Up to seven different methods\footnote{All methods have been adapted from those used to analyse the UVES targets of the Gaia-ESO Survey.} were considered for this analysis, which used equivalent width measurements, such as those from DAOSPEC \citep{2008PASP..120.1332S} and ARES \citep{2007A&A...469..783S} to determine abundances with codes like MOOG \citep{1973PhDT.......180S} and GALA \citep{2013ApJ...766...78M}. Additional methods based on synthetic spectra like SME \citep{1996A&AS..118..595V}, MATISSE \citep{2006MNRAS.370..141R} and Turbospectrum \citep{2012ascl.soft05004P} were employed. 
   
   Each method considered the same 1D-LTE MARCS atmospheric models \citep{MARCS} and line list created for the Gaia-ESO Survey (Heiter et al 2014, in prep). In addition, they used  the same value for rotational velocity (taken from the literature), but micro- and macro-turbulent velocities were determined simultaneously with \feh. Finally, each group analysed the same spectra, which are part of the spectral libraries of \bs s (see \sect{lib}) and cover the wavelength range from 480 to 680~nm.
    
 The final result of \feh\ was obtained by combining the abundances obtained for each iron line individually. We considered the values of only those lines where at least three methods obtained an abundance that agreed within $2\sigma$ with the mean abundance. In addition, NLTE corrections were applied for each line individually using the corrections  of \cite{2012MNRAS.427...50L}. Finally, the \feh\ was calculated from the mean of the selected lines relative to the solar abundance value of \cite{2007SSRv..130..105G}. 
 
We have given special importance in quantifying some of the many different sources of errors that affect the metallicity. Among them we have quantified: (i) the intrinsic scatter of the line-to-line analysis; (ii) the difference obtained in the metallicity after considering the errors associated to \teff\, \logg\ and \vmic \footnote{This is done by  determining the metallicity by fixing \teff, \logg\ and \vmic\ considering their associated uncertainties.}; (iii) the difference obtained from ionised and neutral iron abundances; (iv) the difference obtained from neutral iron abundances under LTE and NLTE.  Details of this analysis can be found in Paper~III.


 
\section{Spectral Library}\label{lib}
Since the Gaia Benchmark  Stars are located in both hemispheres, not all of them are observed with the same instrument. For that reason, the spectra of the Gaia Benchmark  Stars have different properties such as resolution, wavelength coverage and flux calibrations.  The first version of our library includes a collection of  spectra from three instruments:   NARVAL in France and UVES and HARPS in Chile. 

In order to use the Benchmark  Stars as calibrators for a given data set, our spectra must be adapted to the resolution, wavelength range and sampling of these data.  We have developed a software for this purpose which is also presented in this proceeding by \cite{sergi_proceeding}. 
As an application of this software, we transformed our original observations to look like UVES data for the Gaia-ESO survey,  meaning that we resampled and convolved them to the FLAMES-UVES resolution and selected the desired wavelength range. We also determined the radial velocity and shifted the spectra to rest frame.  For the metallicity determination, from the high resolution spectra of our library, we created a data set like  FLAMES-UVES spectrum, but with higher resolution. With our program we are able to edit the format of the original archive spectra and create spectral libraries  with any desired spectral resolution and wavelength coverage (within the limits of the original spectra).  A full documentation on the original spectra and how our tools work is found in Paper~II\footnote{Additional material on the library can be found in the public ftp directory 
{\tt ftp://ftp.obs.u-bordeaux1.fr/pub/jofre/Libraries/Benchmark4GES/} }.

\begin{figure}[ht!]
 \vspace{1cm}
\begin{center}
\includegraphics[scale=0.6]{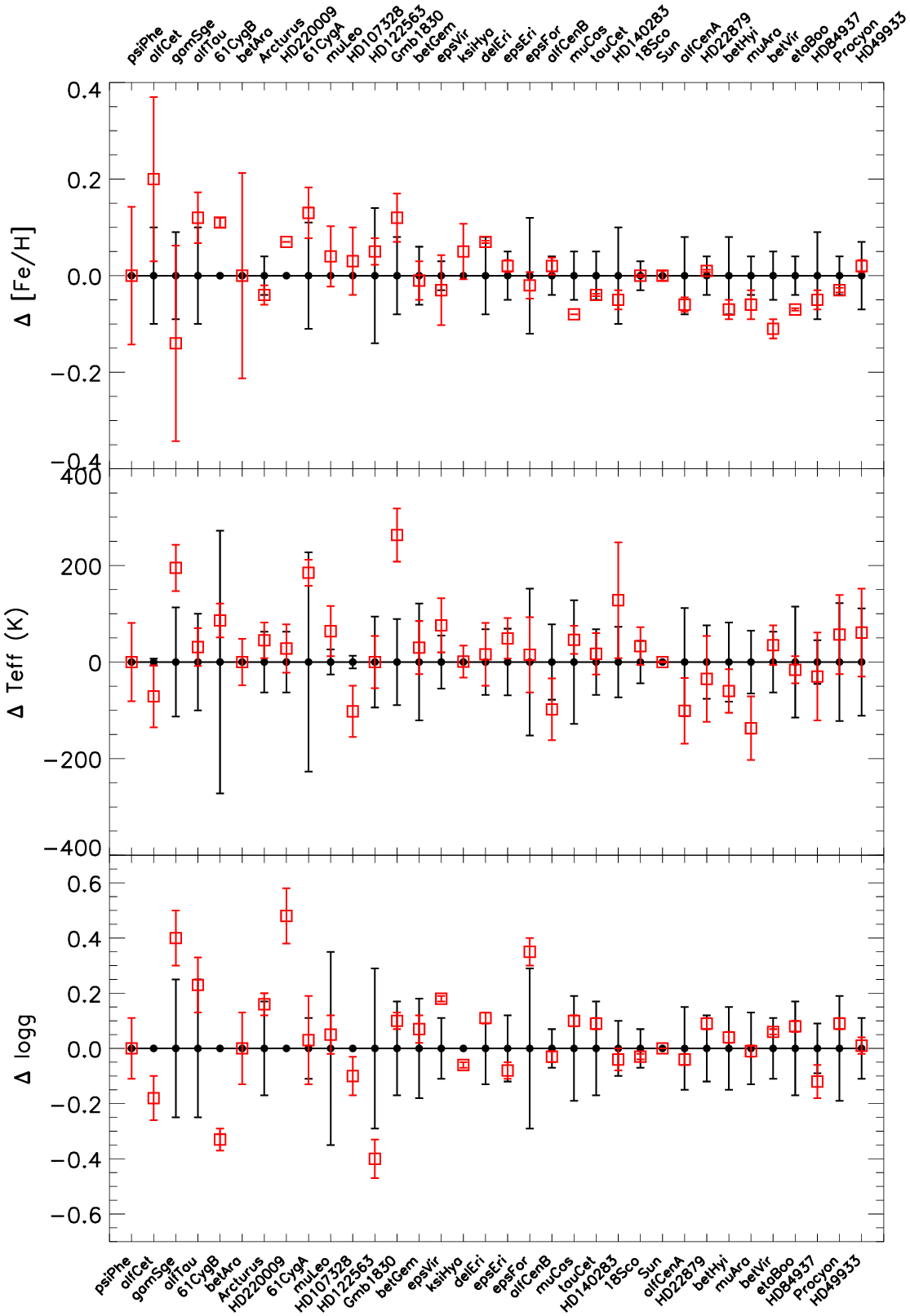}
  \caption{Comparison of our parameters with previous works reported between the years 2000 and 2012 in the literature as retrieved from PASTEL \citep{2010A&A...515A.111S}.  Black error bars indicate the standard deviation in the results of parameters from the literature. Red error bars indicate the mean absolute error of our results (see text). The squares represent the difference between our atmospheric parameters and the average of the literature parameters from PASTEL.  }
  \label{params_plot}
  \end{center}
\end{figure}

 \section{Results}\label{results}

 A comparison of our parameters with the literature can be seen in \fig{params_plot}. The literature value is the averaged one of all references found in the PASTEL catalogue \citep{2010A&A...515A.111S} between the years 2000 and 2012 and the standard deviation of the mean is plotted as black error bars in the figure.  We show the difference between our values and the values from the literature for the effective temperature, surface gravity and metallicity with red squares in three different panels. The red error bars correspond to the mean absolute error in our parameters.

 The large scatter found in the literature reflects the significant differences of the spectroscopic methods. An important result of our work is that we provide homogeneous atmospheric parameters obtained by one method, and where \teff\ and \logg\ are determined independently of spectroscopy.
 
One can see that some of our values differ significantly from the mean literature ones. Some extreme cases are  HD220009, Gmb~1830 and \gamSge, whose direct effective temperatures are very different to the spectroscopic or photometric ones.  At variance, the various measurements for the Sun agree very well, but this may reflect the fact that methods are usually calibrated on this star.  One important reason for the differences seen between the parameters of our work and the literature, is that the common way to determine parameters from the spectra are very model dependent. These models are imperfect. For some cases, the stellar parameters can change significantly with respect to those determined independently from the spectrum.  

  Table~\ref{params} shows the parameters for \teff, \logg\ and \feh\ taking into account the errors mentioned in \sect{params_sect}.  Since we are actively working on defining a consistent and well established set of parameters and associated errors, the values listed in \tab{params} are approximative.  The final values can be found in the corresponding Paper ~I and~III.


 \section{Conclusions}\label{conclusions}
  
We presented a summary of on-going work on FGK Gaia Benchmark  Stars by providing their basic information such as coordinates, magnitudes and stellar parameter ranges in \tab{params}. In addition, we can provide observed spectral libraries that can simulate different data sets. The Gaia-ESO Survey is currently using  this small, but representative sample of FGK reference stars to test the different methods and homogenise their results, showing already advantages of having more well established pillars than the Sun.   Because \teff\ and \logg\ are independent of the spectra, we encourage colleagues working on other spectroscopic surveys to calibrate and link their pipelines with the current Gaia-ESO ones using the Gaia \bs s as well.

The Gaia Benchmark  Stars will be used as a pillar for calibrations of Gaia and other spectroscopic surveys, therefore it is of extreme importance to have a well defined set of atmospheric parameters.   As commented above, spectroscopic and fundamental parameters can differ significantly for some stars.  This reminds us of 
the importance of having standard set of stars with stellar parameters that are independent from spectroscopy, as this helps to make improvements to  spectroscopic methods and models for different kinds of spectral-type stars.
 
 Although our metallicities are scaled to the Gaia-ESO material, such as line list, atmospheric  models and wavelength range, we look forward to  improve our values in a way that it is consistent with methods outside Gaia-ESO. The connection between surveys is crucial to have consistent parameters to understand the  history of the Milky Way.

 \section*{Acknowledgments}
{\small We are pleased to  thank a large group of colleagues who contribute  to this work. They provide  fruitful discussions and helped with selection criteria, preparation of data and the spectral analysis. We acknowledge the contribution of F. Th\'evenin, B. Gustafsson and A. Korn in the selection of targets and determination of stellar parameters.  The production of the first versions of the spectral library  was done with help of  N. Brouillet, T. Jacq and G. Sacco. Finally,  the metallicity determination and its applications in the Gaia-ESO Survey was done together with many colleagues of the Survey,  among them G. Gilmore, S. Randich, M. Bergemann, T. Cantat-Gaudin,   J. Gonzalez Hernandez,  V. Hill, C. Lardo, P. de Laverny, K. Lind, L. Magrini,  T. Masseron, D. Montes,  A. Mucciarelli, T. Nordlander, E. Pancino, A. Recio-Blanco, G. Sacco, R. Smiljanic, J. Sobeck, R. Sordo,  S. Sousa, H. Tabernero,  A. Vallenari, S. Van Eck and C.~C. Worley. P.J. acknowledges support of T. M\"adler in assisting to this workshop and discussing this manuscript. Finally, we thank the referee for useful comments in this proceeding.}

\begin{table}
\hspace{-2.0cm}
\begin{tabular}{c |c c c c c c c}

\hline \hline
star & RA & DEC & spec. type & V &\teff  (K) & \logg   & \feh  \\
\hline
{\bf F dwarfs}  \\
Procyon &07 39 18.119&+05 13 29.96&F5IV-V&0.4&6550$\pm$80&3.99$\pm$ 0.02&  +0.01$\pm$ 0.03\\
HD~49933&06 50 49.832&-00 32 27.17&sdF5& 8.3 & 6640$\pm$90&4.21$\pm$ 0.03& --0.41$\pm$ 0.02 \\
HD~84937&09 48 56.098&+13 44 39.32 &F2V&5.8& 6360$\pm$90&4.11$\pm$ 0.06& --2.03$\pm$ 0.04  \\
\hline
{\bf FGK subgiants} \\
\delEri&03 43 14.901&-09 45 48.21&K1III-IV& 3.5& 5050$\pm$70&3.77$\pm$ 0.02&  +0.06$\pm$ 0.01 \\
HD~140283 &15 43 03.097&-10 56 00.60&sdF3&7.2& 5610$\pm$120&3.67$\pm$ 0.04& --2.36$\pm$ 0.04\\
\epsFor &03 01 37.637&-28 05 29.60&K2VF&5.93& 5120$\pm$80&3.45$\pm$ 0.05& --0.60$\pm$ 0.04\\
\etaBoo &13 54 41.079&+18 23 51.79&G0IV&2.7&6110$\pm$30&3.80$\pm$ 0.02&  +0.32$\pm$ 0.02\\
\betHyi &00 25 45.070&-77 15 15.29&G0V&2.8&5870$\pm$50&3.98$\pm$ 0.02& --0.04$\pm$ 0.03\\
\hline
{\bf Solar-type stars} \\
\alfCen A &14 39 36.494&-60 50 02.37&G2V&0.1&5850$\pm$70&4.31$\pm$ 0.02&  +0.26$\pm$ 0.03  \\
HD~22879 &03 40 22.064&-03 13 01.12&F9V&6.7&5870$\pm$90&4.23$\pm$ 0.02& --0.86$\pm$ 0.01 \\
Sun & --- & --- &G2V&-26.8&5777&4.44&  +0.01$\pm$ 0.01 \\
\muCas &01 08 16.395&+54 55 13.23&G5Vb&5.2& 5310$\pm$30&4.41$\pm$ 0.02& --0.81$\pm$ 0.01\\
\tauCet &01 44 04.083&-15 56 14.93&G8.5V&3.5&  5330$\pm$40&4.44$\pm$ 0.02& --0.49$\pm$ 0.02 \\
\alfCen B &14 39 35.063&-60 50 15.10&K1V&1.4&  5260$\pm$60&4.54$\pm$ 0.02&  +0.22$\pm$ 0.02 \\
18~Sco &16 15 37.269&-08 22 09.99&G2Va&5.5&5750$\pm$40&4.43$\pm$ 0.01&  +0.03$\pm$ 0.01\\
\muAra &17 44 08.701&-51 50 02.59&G3IV-V&5.1&5900$\pm$70&4.27$\pm$ 0.02&  +0.35$\pm$ 0.04 \\
\betVir &11 50 41.718&+01 45 52.99&F9V&3.6& 6080$\pm$40&4.08$\pm$ 0.01&  +0.24$\pm$ 0.03 \\
\hline 
{\bf Red Giants} \\
Arcturus &14 15 39.672&+19 10 56.67&K1.5III&-0.1&4250$\pm$40&1.59$\pm$ 0.04& --0.52$\pm$ 0.04\\
HD~122563 &14 02 31.845&+09 41 09.95&F8IV&6.2& 4590$\pm$50&1.61$\pm$ 0.07& --2.64$\pm$ 0.08 \\
\muLeo &09 52 45.817&+26 00 25.03&K2III&3.9& 4470$\pm$50&2.50$\pm$ 0.07&  +0.25$\pm$ 0.07\\
\betGem &07 45 18.950&+28 01 34.32&K0IIIb&1.1&4860$\pm$60&2.88$\pm$ 0.05&  +0.13$\pm$ 0.09\\
\epsVir &13 02 10.598&+10 57 32.94&G8III&2.8& 4980$\pm$60&2.77$\pm$ 0.01&  +0.15$\pm$ 0.08 \\
\ksiHya &11 33 00.115&-31 51 27.44&G7III&3.5& 5040$\pm$30&2.87$\pm$ 0.01&  +0.16$\pm$ 0.11 \\
\alfTau &04 35 55.239&+16 30 33.49&K5III&0.9& 3930$\pm$40&1.22$\pm$ 0.10& --0.37$\pm$ 0.56  \\
\psiPhe &01 53 38.741&-46 18 09.60&M4III&4.4&3470$\pm$80&0.62$\pm$ 0.11& --1.24$\pm$ 0.14 \\
\gamSge &19 58 45.429&+19 29 31.73&M0III&3.5& 3810$\pm$50&1.05$\pm$ 0.10& --0.17$\pm$ 0.20 \\
\alfCet &03 02 16.773	& +04 05 23.06&M1.5IIIa&2.5& 3800$\pm$60&0.91$\pm$ 0.08& --0.45$\pm$ 0.17\\
\betAra &17 25 17.988	 &-55 31 47.59&K3Ib-II&2.8&4170$\pm$50&1.01$\pm$ 0.13& --0.05$\pm$ 0.21 \\
HD~220009&23 20 20.583	 &+05 22 52.70& K2III&5.0&  4280$\pm$50&1.43$\pm$ 0.10& --0.74$\pm$ 0.06 \\
HD~107328 &12 20 20.981	& +03 18 45.26&K0IIIb&5.0& 4500$\pm$50&2.11$\pm$ 0.07& --0.33$\pm$ 0.07\\
\hline
{\bf K dwarfs} \\
\epsEri &03 32 55.845	& -09 27 29.73&K2Vk&3.7& 5050$\pm$40&4.60$\pm$ 0.03& --0.09$\pm$ 0.02 \\
Gmb~1830&11 52 58.769	& +37 43 07.23&	G8Vp&6.5& 4830$\pm$60&4.60$\pm$ 0.03& --1.46$\pm$ 0.05 \\
\cyg A &21 06 53.952	& +38 44 57.99&K5V&5.2& 4340$\pm$30&4.43$\pm$ 0.16& --0.33$\pm$ 0.05 \\
\cyg B &21 06 55.264	& +38 44 31.40&K7V&6.0&  4050$\pm$40&4.53$\pm$ 0.04& --0.38$\pm$0.03\\
\hline \hline

\end{tabular}
\caption{General information and atmospheric parameters of the Gaia FGK \bs s. The first three columns indicate the name of the star and its coordinates. The fourth column lists the spectral type, while the fifth one denotes the V magnitude.  The last three columns indicate the temperature, surface gravity and metallicity range, respectively.}              
\label{params}
\end{table}



\begin{thebibliography}{}
%

\bibitem[Blanco-Cuaresma et al.(2013)]{sergi_proceeding} Blanco-Cuaresma, S., Soubiran, 
C., Jofr\'e, P. \&. Heiter, U., 2013, IWSSL Proceedings 

\bibitem[Gilmore et al.(2012)]{2012Msngr.147...25G} Gilmore, G., Randich, 
S., Asplund, M., et al.\ 2012, The Messenger, 147, 25 

\bibitem[Grevesse et al.(2007)]{2007SSRv..130..105G} Grevesse, N., Asplund, 
M., \& Sauval, A.~J.\ 2007, SSRv, 130, 105 

\bibitem[Gustafsson et 
al.(2008)]{MARCS} Gustafsson, B., Edvardsson, B., Eriksson, K., et al.\ 2008, A\&A, 486, 951 


\bibitem[Jofr\'e et al.(2013)]{2013arXiv1309.1099J} Jofr\'e, P., Heiter, U., 
Soubiran, C., et al.\ 2013, arXiv:1309.1099 


\bibitem[Lind et al.(2012)]{2012MNRAS.427...50L} Lind, K., Bergemann, M., 
\& Asplund, M.\ 2012, MNRAS, 427, 50 

\bibitem[Mucciarelli et al.(2013)]{2013ApJ...766...78M} Mucciarelli, A., 
Pancino, E., Lovisi, L., Ferraro, F.~R., 
\& Lapenna, E.\ 2013, ApJ, 766, 78 

\bibitem[Plez(2012)]{2012ascl.soft05004P} Plez, B.\ 2012, Astrophysics  Source Code Library, 5004 


\bibitem[Recio-Blanco et al.(2006)]{2006MNRAS.370..141R} Recio-Blanco, A., Bijaoui, A., \& de Laverny, P.\ 2006, MNRAS, 370, 141 

\bibitem[Sneden(1973)]{1973PhDT.......180S} Sneden, C.~A.\ 1973, 
Ph.D.~Thesis,  

\bibitem[Soubiran et al.(2010)]{2010A&A...515A.111S} Soubiran, C., Le Campion, J.-F., Cayrel de Strobel, G., \& Caillo, A.\ 2010, A\&A, 515, A111 

\bibitem[Sousa et  al.(2007)]{2007A&A...469..783S} Sousa, S.~G., Santos, N.~C., Israelian, et al.\ 2007, A\&A, 469, 783 

\bibitem[Stetson \& Pancino(2008)]{2008PASP..120.1332S} Stetson, P.~B., \& Pancino, E.\ 2008, PASP, 120, 1332 


\bibitem[Valenti \& Piskunov(1996)]{1996A&AS..118..595V} Valenti, J.~A., \& Piskunov, N.\ 1996, A\&AS, 118, 595 





























\end{thebibliography}
\end{document}